\documentclass[11pt,a4paper]{article}

\usepackage[utf8]{inputenc}
\usepackage{array}
\usepackage{bibentry}  
\nobibliography*       
\usepackage{footmisc}       
\usepackage{geometry}
\usepackage{longtable}
\usepackage{pdflscape}
\usepackage{todonotes}
\usepackage{xspace} 
\usepackage{tikz}
\usetikzlibrary{arrows.meta}
\usetikzlibrary{shapes.geometric}
\usepackage{setspace}
\usepackage{hyperref}
\usepackage{etoolbox}
\appto\UrlBreaks{\do\-}

\hypersetup{
	pdfinfo={
		Title = {Collections of Android Applications},
		Author = {Franz-Xaver Geiger, Ivano Malavolta},
		Subject={Literature review},
		Keywords={Android, Mining Software Repositories, Dataset, Source Code},
	},
	hidelinks,
	linktoc=all
}

\newcolumntype{R}[1]{>{\raggedright\arraybackslash}p{#1}}

\newcommand{\name}[1]{\emph{#1}\xspace}

\newcommand{\manifest}{\texttt{AndroidManifest.xml}\xspace}
\newcommand{\play}{\name{Google Play}}
\newcommand{\github}{\name{GitHub}}
\newcommand{\fdroid}{\name{F-Droid}}
\newcommand{\eg}{e.g.\ }
\newcommand{\etal}{et al.\xspace}
\newcommand{\ie}{i.e.,\ }
\newcommand{\etc}{etc.\xspace}
\newcommand{\cf}{cf.\ }
\newcommand{\filename}[1]{\texttt{#1}}

\title{
    \huge Datasets of Android Applications: a Literature Review
}
\author{
    Franz-Xaver Geiger, Ivano Malavolta \\
    \small\itshape f.geiger@student.vu.nl, i.malavolta@vu.nl \\
	\small Vrije Universiteit Amsterdam, The Netherlands
}

\begin{document}

\maketitle

\abstract{
Mobile phones and tablets have become the most widely used computing devices, with a large predominance of the Android platform.
As a natural evolution, the development of Android applications has surged and has become a major field of study,
with research efforts ranging from energy efficiency, to code smells, performance, maintainability, security, etc. 
These kind of challenges ask for dedicated solutions, tools, and datasets.

This survey identifies and reviews 31 existing datasets of Android applications and classifies each of them according to key features, such as the total number of apps it contains, whether the commit history of the apps is available, whether it focusses on the source code or on the executable binaries of the apps, the sources used for building the dataset, etc.

This study can benefit both the experienced and the novice researcher interested on doing research on Android apps, which can use the results of our study as a map for identifying the most suitable datasets for their research objectives.
}

\section{Introduction}\label{sec:intro}

Mobile phones and tablets have become the most widely used computing devices.
Consequently, development of mobile applications has surged and become a major field of study.
Additionally, mobile platforms bring their particular set of constraints.
For instance, energy on small devices is a scarcity and power management is paramount.
Privacy of users and software security are other highly studied topics.
These kind of challenges ask for dedicated solutions, tools, and datasets.

This survey reviews Datasets of Android applications.
However, not all research needs the same set of data.
Martin \etal\ provide an extensive survey of studies and datasets of app store analysis for various platforms~\cite{martin_survey_2017}.
They identified seven key subfields:
\emph{API Analysis},
\emph{Feature Analysis},
\emph{Release Engineering},
\emph{Review Analysis},
\emph{Security},
\emph{Store Ecosystem},
\emph{Size and Effort Prediction}, and
\emph{Closely Related Work} (among which is \emph{Mining Tools}).
Research may be interested in technical attributes such as API usage or platform version, as well as non-technical attributes, \eg reviews, number of downloads, \etc\@
For dynamic analysis of applications, executable artifacts are necessary.
Bytecode from APKs can be decompiled to learn information about data flow and other code metrics.
To analyze apps for programming practices and project management, source code and data from source code management programs such as version control and bug trackers is helpful.
The latter category of information is not readily available for the vast number of proprietary applications.
Studies that need this kind of data need to rely on open-source Android apps.

Therefore, we review existing literature for various characteristics which may facilitate different sub-fields and studies.
This survey thus focuses on these main traits of datasets of Android applications:
\begin{itemize}
	\item Does the dataset facilitate access to source code of applications?
	\item Is the source code available in version control (\eg Git)?
	\item Are installable APKs included?
	\item Does the dataset link to app stores where additional information (such as ratings and reviews) are accessible?
\end{itemize}

This literature survey is structured as follows:
First, in Section~\ref{sec:methods}, we explain the iterative literature review process from keyword search and snowballing to a concise view of important information in a table.
Following that we review and summarize datasets and studies resulting from the search process (Section~\ref{ch:related}).
Learnings from the review results are detailed in Section~\ref{sec:reflection} where we argue that too few datasets include access to source code and those that link source code contain too few applications.
Finally, we conclude this survey in Section~\ref{sec:related:conclusion}.

\section{Literature Review Process}\label{sec:methods}

\begin{figure}[!bt]
	\centering
	\tikzset{
    node distance = 1.5cm,
    > = {Latex [width = 5, length = 5]},
    rounded corners,
    box/.style = {
        draw,
        minimum width = .75\textwidth,
        minimum height = 2em,
        text width = .66\linewidth,
        align = left,
    },
	label of/.style 2 args = {
        rectangle,
        fill = #2!15,
        anchor = east,
        at = (#1.east),
        font = \scriptsize,
        minimum width = 5em,
        minimum height = 1.8em,
        xshift = -.1em,
        align = center,
	},
    arrow label/.style = {
        draw = none,
        minimum width = 0pt,
        minimum height = 0pt,
        align = center,
        font = \tiny,
    },
    decision/.style = {
        draw,
        diamond,
        aspect = 2,
    },
}
\begin{tikzpicture}

    \node (keyword) [box]
        { Search by keywords };
    \node (title-filter)
        [box, below of = keyword]
        { Filter relevant publications by title };
    \node (abstract-filter)
        [box, below of = title-filter]
        { Filter relevant publications by abstract };
    \node (continue-decision)
        [decision, below of = abstract-filter, yshift = 3mm]
        {$n$};
    \node (snowball-cited)
        [box, below of = continue-decision]
        { List publications cited by search results};
    \node (snowball-citing)
        [box, below of = snowball-cited]
        { List publications citing search results };
    \node (summarize)
        [box, below of = snowball-citing]
        { Summarize each publication };
    \node (table)
        [box, below of = summarize]
        { Describe datasets in table };
    \node (categories)
        [box, below of = table]
        { Categorize collections of Android apps };

    \node [label of = {keyword}{blue}] { Phase 1 };
    \node [label of = {title-filter}{red}] { Phase 2.1 };
    \node [label of = {abstract-filter}{red}] { Phase 2.2 };
    \node [label of = {snowball-cited}{orange}] { Phase 3.1 };
    \node [label of = {snowball-citing}{orange}] { Phase 3.2 };
    \node [label of = {summarize}{green}] { Phase 4.1 };
    \node [label of = {table}{green}] { Phase 4.2 };
    \node [label of = {categories}{green}] { Phase 4.3 };

    \draw [->] (keyword) to (title-filter);
    \draw [->] (title-filter) to (abstract-filter);
    \draw [->] (abstract-filter) to (continue-decision);
    \draw [->] (continue-decision) -- (snowball-cited) node [arrow label, near start, left] {$n > 0$};
    \draw [->] (snowball-cited) to (snowball-citing);
    \draw [->] (snowball-citing.west) -| ++(-1cm,2.5cm) |- (title-filter.west);
    \draw [->] (continue-decision.east) -| ++(5.5cm,-2.5cm) node [arrow label, near start, above] {$n = 0$} |- (summarize);
    \draw [->] (summarize) to (table);
    \draw [->] (table) to (categories);

\end{tikzpicture}
	\caption{Literature review process}\label{fig:search-process}
\end{figure}
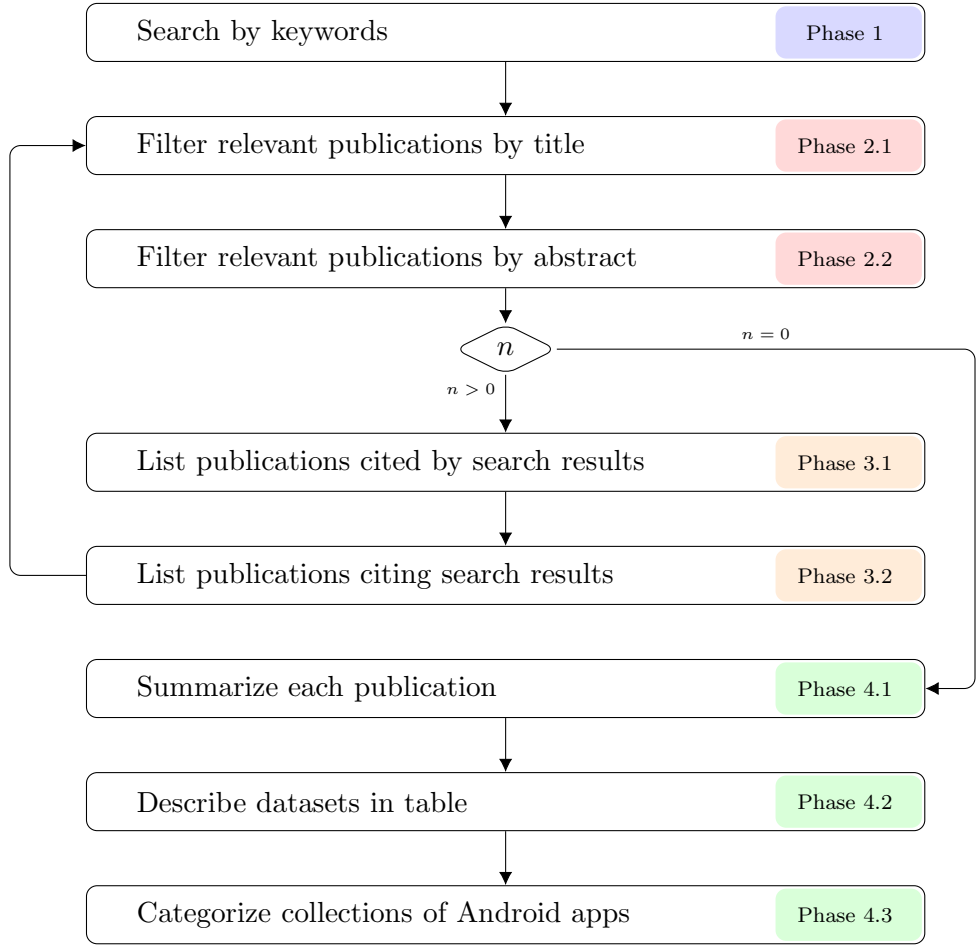

Literature presented in this review was collected with a combination of keyword search and snowballing, \ie walking the graph of references in both directions.
All queries were ran against the \name{Google Scholar} database in winter 2017/18.
The review is concerned with datasets of Android applications in general but also more specifically with datasets that allow access to source code of apps.
Figure~\ref{fig:search-process} shows the iterative search process which followed four steps, repeating phase~2 and phase~3 until search results were exhausted.
The four steps are
(1) an initial keyword search,
(2) filtering of relevant publications by title and abstract,
(3) finding candidate publications by following the graph of citations from new search results to both citing and cited articles, and finally,
(4) summarizing all found relevant publications in textual and tabular form.

\paragraph{Phase 1: Keyword search}
Initially we searched for \emph{``Android app dataset''}, and \emph{``Android app collection''}
\emph{``Android app mining''}.
The search results were complemented by replacing the keyword \emph{app} with \emph{application} in each search term.
Filtering the search results for relevant publications showed one major group of publications around the topic of Android application security.
These publications are largely centered around \emph{AndroZoo}~\cite{allix_androzoo:_2016} and the \emph{Android Malware Genome Project}~\cite{zheng_droid_2013}.
To broaden the search scope and find datasets including source code, the search terms \emph{``android app'' ``source code'' repository dataset} were included.

\paragraph{Phase~2.1: Title filter}
The search results at this point were filtered to exclude publications that are obviously out of scope for this review by looking at their titles.

\paragraph{Phase~2.2: Abstract filter}
After reducing the scope by title, we read through abstracts of all search results and filtered those out which do not create a dataset of Android applications.
We looked for indicators, that the paper actually gathers data on Android applications or uses a dataset to study Android apps.
Only in the former case did we include the publication in my set of relevant work.
In the latter case, we did not deem the paper itself relevant to my review but included it in the snowballing phase to find further links to existing datasets.

If the filter of Phase~2 yielded new results, Phase~3 was revisited.
Otherwise, the collection phases were concluded and we would continue with Phase~4.

\paragraph{Phase~3.1: Cited publications}
In a next step, we followed links from new papers collected so far to find relevant publications which are cited by them.
This allowed to find previous works which the authors of already identified publications deem relevant to the subject.

\paragraph{Phase~3.2: Citing publications}
We also searched for publications which refer to papers already in my set of relevant works.
While looking at cited publications allows to glance into the past of related literature, searching for articles which cite already known papers gives information about the future from the time of these papers.

This new list of candidate articles was then fed into the filtering process (Phase~2).

\paragraph{Phase~4.1: Summaries}
Phase~4 started after the data collection process was complete with 28 relevant publications and repeating phases~2 and~3 did not return any relevant new publications.
We read all search results and briefly summarized them (\cf Section~\ref{ch:related}).

\paragraph{Phase~4.2: Tabular data overview}
Data from these summaries was then processed into a table (\cf Appendix~\ref{appendix:results}).

\paragraph{Phase~4.3: Categories of datasets}
Finally, we categorized datasets of Android apps which we found in the literature into
(1) datasets which use data from app markets (\cf Section~\ref{sec:related:market-data}),
(2) datasets providing executable APKs (\cf Section~\ref{sec:related:apk}), and
(3) datasets with access to source code based on \name{F-Droid} (\cf Section~\ref{sec:related:fdroid}).
\section{Datasets of Android Apps in the Literature}
\label{ch:related}


In the following sections we describe literature included in this survey which provides access to different levels of information.
An overview of all included publications with relevant traits in tabular form can be found in Appendix~\ref{appendix:results}.

The information collected about Android applications may contain metadata from app stores, such as \play (Section~\ref{sec:related:market-data}).
In Section~\ref{sec:related:apk} we list previous work that contains executable Android application packages.
A directory of open-source Android apps is \fdroid. Datasets that provide access to source code and commit history are often based on it (Section~\ref{sec:related:fdroid}).
In Section~\ref{sec:reflection}, we reflect on the findings of this literature study and propose future directions to improve the state of Android app datasets.
Finally, Section~\ref{sec:related:conclusion} is a summary of the common characteristics and problems of reviewed datasets.

\subsection{Datasets of Market Data}\label{sec:related:market-data}

Many interesting insights can be learned from data on application markets and aggregations of that data.
Official app stores, such as \play\footnote{\url{https://play.google.com/store/apps}} contain several million apps for the Android platform.
Official and inofficial market places host executables and metadata generated by developers and users for each application.
Data from \play can only be accessed through the public web interface and an undocumented API used by Android smartphones to manage app installations.
Commercial databases exist that mirror metatada from \play and other app markets and sell access to this information (\eg \href{https://appannie.com}{appannie.com}, \href{https://appbrain.com}{appbrain.com} and \href{https://appszoom.com}{appszoom.com}~\cite{krutz_dataset_2015}).
Some of these commercial databases contain comprehensive metadata of millions of apps but they lack links to other resources, such as source code or executable artifacts.

Data from market places is widely used despite the difficulties to access it. 
Petsas \etal~\cite{petsas_rise_2013} monitored different Android app stores with a focus on popularity, pricing, and revenue of apps.
They directly scraped information from the web interfaces of the market places in their study.
Their findings indicate that 10 percent of the apps account for 70 to 90 percent of total downloads
and that popularity of paid apps follows a power law distribution.

Another valuable data point from market places are user reviews.
Malavolta \etal~\cite{malavolta_hybrid_2015} investigated users' perception of hybrid apps by studying 11,917 free apps and their metadata from \play.
They answered questions from both developers' and users' perspective by combining user reviews and technical aspects in their study.
In the data collection process they selected sample apps from the most popular apps of each category in \play.
Grano \etal~\cite{grano_android_2017} also studied user reviews albeit from a different source:
They built a dataset of 288,065 user reviews for 395 applications sourced from \fdroid.
The dataset includes information from \play, as well as results from static analysis of the application packages.
They labeled reviews with automated classifiers.
Other studies use
permissions of apps~\cite{moonsamy_mining_2014},
API usage of apps~\cite{aafer_droidapiminer:_2013,linares-vasquez_mining_2014},
descriptions~\cite{gorla_checking_2014},
or times of updates of apps~\cite{mcilroy_fresh_2016}
from \play.

This wide field of research on data from application markets shows that app metadata, user reviews, and app binaries offer insights and are worth investigating.
However, access restrictions and instable APIs limit the use of app stores to the research community.

\subsection{Datasets of Executables}\label{sec:related:apk}

On the other hand, \name{AndroZoo} is an ongoing effort to gather executable Android applications from as many sources as possible and make them available for analysis.
Allix \etal~\cite{allix_androzoo:_2016} created crawlers for several app stores to collect a comprehensive and up-to-date sample of executable Android app packages --- \emph{AndroZoo}.
The crawlers are customized for each app store to collect as many apps as possible.
Simultaneously, the authors took measures to minimizing the load on market places they crawl to avoid losing access and jeopardizing long-term integrity of the dataset.
The sources from which \emph{AndroZoo} draws include major market places \play, \emph{Anzhi}, and \emph{AppChine}, as well as smaller directories \emph{1mobile}, \emph{AnGeeks}, \emph{Slideme}, \emph{ProAndroid}, \emph{HiApk}, and \emph{F-Droid}.
The applications from these app stores were complemented with additional artifacts from peer-to-peer distributed torrents and the \emph{Android Malware Genome Project}~\cite{zhou_dissecting_2012}.
The procedure to download candidate apps is performed by dedicated crawlers for each source and includes a unique identifier and a checksum of the file for deduplication.
Most crawlers are based on the scrapy framework.
However, Allix \etal created a special software to overcome restrictions of \play, \eg an undocumented API, rate limits, and the need of an Android device.
A central dispatcher spreads the work load to download agents in several locations and over different protocols.
With this setup it was possible to eliminate the backlog of old applications.
Subsequently, fewer agents were necessary to keep up with new additions to \play.
A web service is tasked with organizing and storing received APKs.
This unit also handles authentication for downloads of the dataset and publicly displays statistics.
When creating \emph{AndroZoo}, Allix \etal encountered several data collection challenges.
They list unexpected downtime of markets, HTML instability, monitoring of crawlers, protocol changes, and information loss.
Overall, the authors were able to collect more than three million Android applications initially.
The current count is more than five million \cite{li_mining_2017}.
The majority of these apps stems from \play, \emph{Anzhi}, and \emph{AppChine}, with the other market places contributing a much lower number.
The dataset is available for download for the research community as a regularly updated list of APKs.
This list contains SHA256 hashes as identifiers and additional metadata, such as compilation date, malware status, package name, version, \etc\@
Individual apps can be downloaded with the SHA256 hash as index.
One defining feature of \emph{AndroZoo} is, that all apps in the dataset are tested for malware by over 60 security products hosted by \emph{VirusTotal}.
Allix \etal report that 22 percent of apps in \play are flagged as malware by at least one product while 50 percent or more are found to be malware in the two major Chinese market places.
When counting APKs which at least ten security products recognize as malware, this number drops to around 1 percent of detected malware in \play and 33 percent and 17 percent in \emph{Anzhi} and \emph{AppChine} respectively.
All samples of the \emph{Android Malware Genome Project} are successfully recognized by at least 10 antivirus products.
The dataset lends itself to security research since metadata of all samples contains the malware detection status.
Examples of such research based on \emph{AndroZoo} are \cite{allix_empirical_2016,allix_are_2015,allix_forensic_2014,hurier_lack_2016}.
Other uses leverage the fact that the dataset contains several version for many apps~\cite{hecht_tracking_2015} and the availability of compiled bytecode~\cite{li_iccta:_2015}.
\emph{AndroZoo} also contains many Android applications which are not marketed in \play.
This facilitates analysis of marketed and non-marketed apps~\cite{nayebi_analysis_2016}.
Limitations of \emph{AndroZoo} mostly stem from the fragility of the data collection process.
Collecting was not continuous but rather resumed irregularly, if issues occurred.
Additionally, app some market maintainers have blocked crawlers and thus caused outages and incomplete sets of data.


Another dataset of Android applications is the \emph{Android Malware Genome Project}~\cite{zhou_dissecting_2012}.
Zhou and Jiang collect samples of malicious Android apps from August 2010 to October 2011 to advance understanding of malware on mobile platforms.
They present a dataset of 1,260 apps in 49 different malware categories.
Furthermore, the authors analyze and characterize the collected malware samples to trace behavior and major outbreaks of certain types.
Zhou and Jiang report that most of the samples are repackaged versions of legitimate applications containing malicious payload.
Another vector for infecting Android devices are update attacks and drive-by downloads.
Types of malware include root-level exploits, botnet clients, incurring costs through calling or messaging to premium-rate numbers, and harvesting of users' information.
In their evolution-based study, Zhou and Jiang describe how Android malware rapidly evolves.
Thus, malware authors are able to keep ahead of existing anti-malware solutions through application of sophisticated obfuscation and evasion techniques.

The project allows studying of generations and classes of malware but does not link these artifacts with source code or version control data.
The authors stopped sharing their data after graduation in 2015.

Recently, Meng \etal~\cite{meng_androvault:_2017} published	\name{AndroVault}, a knowledge graph of information on over five million Android apps.
Since 2013 applications have been crawled from 33 different sources including \play and \fdroid.
The tool computes 36 attributes for each app based on downloaded APKs and descriptions.
Resulting data from static and dynamic analysis is combined in a knowledge graph with fast access.
Entities in this knowledge graph are heuristically clustered and correlated by attributes.
This facilitates easier selection and sampling of relevant apps by certain traits to research specific kind of Android applications.
\name{AndroVault} has already proven a useful dataset for research such as malware detection.


One large user of datasets of Android application packages is the security research community, \eg for evaluation of malware detection systems~\cite{arp_drebin:_2014,zheng_droid_2013,lindorfer_andrubis1000000_2014}.
Malware detection necessarily needs to work on compiled artifacts because that is the form in which it is installed on devices and for which detection is possible.
Datasets of executables are therefore well suited for studying malicious software and training detection systems.
Android application packages are not a substitute for source code and project management data, such as issue trackers and code review.

\subsection{Datasets Based on \fdroid}\label{sec:related:fdroid}

So far all described datasets rely on \play or similar market places as seeds.
This limits the available types of information to market metadata, executable packages and what can be statically or dynamically inferred from the APK files.
In order to enable research that relies on access to source code, data from application markets needs to be linked to additional information.
One data source that provides access to source files is \fdroid:\footnote{\url{https://f-droid.org}}
a directory of open-source Android applications.
All apps listed in this directory are compiled from source and code repositories are publicly linked.

In 2013, Minelli and Lanza~\cite{minelli_software_2013} analyzed Android apps from \fdroid and reported notable findings, such as little use of inheritance and heavy reliance on external APIs.
Freiling \etal~\cite{freiling_empirical_2015} use 240 randomly selected apps from \fdroid to evaluate obfuscation transformations.

Bao \etal~\cite{bao_how_2016} collected 468 commits from 154 \github repositories of Android apps starting from 1,273 apps on \fdroid.
They categorized energy-aware commits in six buckets, corresponding to common power management techniques applied by developers.
They found that types of power management related changes differ between Android apps of different app store categories.

Lamba \etal~\cite{lamba_pravaaha:_2015} extensively describe \fdroid and used 1,120 apps from the app directory to analyze software use for Android applications.
They downloaded the latest version of the source code of all collected apps and ran their analysis on 87,478 Java files with 17.2 million lines of code.
Corral and Fronza~\cite{corral_better_2015} manually combined data from \fdroid, \play, and any available source code repository for 100 apps to compare source code quality with market success.
They report that source code quality has a marginal impact on market success.

Nayebi \etal~\cite{nayebi_analysis_2016} analyzed 1,844 applications from \fdroid and found 69 apps that matched their search criteria.
They linked this data to \github repositories and \play listings for further analysis of release cycles.
``A Dataset of Open-Source Android Applications''~\cite{krutz_dataset_2015} was similarly generated with \fdroid as starting point.
The dataset contains 1,179 entries and links to source code repositories and information gleaned from static analysis of binary artifacts.
It additionally contains version control information, such as commit messages and authorship.
Unfortunately, the website hosting the dataset seems to be defunct.

For a follow-up study, Krutz \etal~\cite{krutz_who_2017} extended this data for detailed analysis of app permissions.
They searched \fdroid for applications with source repositories on \github to find out how and by whom permissions of applications are modified.
To that end, they traced changes to Android manifest files through commit history and analyzed traits of developers who perform these changes.

Das \etal~\cite{das_quantitative_2016} seeded their dataset from various sources in order to achieve wider coverage of available apps.
Next to \fdroid, they also included open-source applications listed on Wikipedia and they searched for links from \filename{Readme} files of \github repositories to \play pages.
In total they found 2,443 open-source Android apps with source code on \github.
Access to version control data allowed them to investigate performance related commits by looking at commit messages stored in \name{Git}.
In summary, their dataset not only contains links to \fdroid with executable APKs, but also references to source code on \github and additional metadata on \play.

Tufano \etal~\cite{tufano_when_2015} manually analyzed 9,164 commits from \name{Git} repositories to investigate how bad programming practices are introduced.
Android app source code is one of the three fields they study.
Their dataset includes 70 apps sourced from \fdroid.
Similarly, Stojkovski~\cite{stojkovski_thresholds_2017} created a dataset of 865 Android applications sourced from \fdroid to study software quality metrics. 
Stojkovski also considered \name{Sourceforge} but did not use it for lack of automated access to Android applications.
As mentioned above, Grano \etal~\cite{grano_android_2017} mined user reviews from \play for a list of apps from \fdroid.
The generated dataset contains 395 apps in around 600 versions.

By resorting to \fdroid as source of Android applications, researchers utilize links to \play and especially to source code repositories.
\fdroid only lists open-source apps and providing source code is inherent to the platform.
This allows researchers to use source code in their analysis and even version control data, such as commit messages and contents.

A drawback of using \fdroid over other market places is, that it only contains 2,697 applications\footnote{As of March 12, 2018} and excludes apps which are not freely licensed.
The number of apps listed on \fdroid is orders of magnitude smaller than on closed source market places, foremost \play.
However, what \fdroid lacks in numbers is compensated by the links to source code repositories with version controlled source code, change reviews, and bug trackers.
\fdroid therefore is a valuable source for mining Android apps.

\subsection{Datasets of Source Code without Reliance on \fdroid}\label{sec:related:source-code}

However, access to source code of Android apps does not need to be restricted to applications in \fdroid.
Linares-Vasquez \etal~\cite{linares-vasquez_how_2015} try a different approach by directly searching \github repositories labeled as containing Java files for \manifest files.
These manifest files are mandatory for and unique to Android apps.
Therefore, they are a good search criteria to identify source code repositories containing code of Android applications.
Linares-Vasquez \etal\ found 16,333 repositories with code for Android apps which is a much higher number than the number of apps available on \fdroid.

Geiger \etal~\cite{geiger_graph-based_2018} use the same idea to initially search for manifest files and construct \name{AndroidTimeMachine}, a graph database of 8,431 Android apps which are both accessible on \github\ and \play.
Their dataset links data from \play\ pages and \github\ repositories and includes metadata of all commits in one \name{Neo4j} graph.

\section{Reflections}\label{sec:reflection}

This literature review found several datasets of Android applications.
They collect and provide executables, market and distribution data, source code, and even analysis results in various forms of detail.

\paragraph{App store data}
One common problem faced by many data datasets is the lack of documented access to data from app stores, especially \play.
Google does not provide a public API and other market places actively block crawlers from collecting data.
Tools do exist to gather app store data from many sources but they heavily rely on regular maintenance and updates to keep working.
Future work could include creating one dataset with comprehensive access to market place data to facilitate research of Android applications.

\paragraph{Updated data}
Researchers have poured a lot of work into creating diverse datasets of Android apps.
Information in these app datasets is capable of shedding light on interesting questions in the field of Android research.
Unfortunately, many of these datasets have not received updates in years.
Information in these datasets turned stale.
Researches facing an ever changing environment of application development cannot rely on these old datasets to perform current research.
This leads to a gap in possible research since newer Android app datasets may not include similar information necessary to answer some research questions.
Future efforts should be directed to update existing datasets and set up new datasets in such a way that they are easier to maintain and kept up-to-date.
Releasing tooling to create a dataset is already a step in the right direction.
Regularly performing the data collection process and making the results available in a versioned format or a timeline should be the next step.

\paragraph{Accessibility of data}
Worse than the problem of outdated data is inaccessible data.
Many datasets of Android applications have not been released publicly or authors stopped sharing them after some time.
It is unfortunate to see that potentially useful data is not shared with the research community.
Instead of re-creating datasets from scratch, building upon previous work and complementing existing data would benefit authors of both old and new publications.
Therefore, researchers should make sure they share data in widely accessible formats and on open platforms to be independent of individual maintenance.
Also including permanent links to data could help make data more easily accessible years after publication.

\paragraph{Source code}
Previous studies and datasets provide different levels of access to data of Android applications.
However, none of the datasets combines all potential data.
Martin \etal~\cite{martin_survey_2017} also highlight a key shortcoming of the literature in its current state:
There are few mining tools and datasets which combine source code with application metadata from app stores and development tools for large sets of apps.
One tool that combines access to all sources mentioned above is \name{CALAPPA}.
To ease access to app market data, Avdiienko \etal~\cite{avdiienko_calappa:_2016} developed a toolchain for mining Android apps.
It has modules for data retrieval from various sources.
This design allows Avdiienko \etal\ to combine app metadata, user reviews, executables, and source code where applicable.
Modules include crawlers and metadata analysis as well as static program analysis and post-processing.
\name{CALAPPA} can retrieve source code for Android apps limited to those listed on \fdroid but does not seem to be publicly available.
Some datasets have increased the number of Android applications for which source code is available.
Unfortunately, this number is still low and the sample of apps is likely biased.
Finding additional means to get access to source code should be on the agenda for future work.

\paragraph{Combining existing data}
Finally, future research could benefit more from existing datasets, if the information contained in them was relatable to information in other datasets.
Various efforts have been undertaken to gather, process, and present relevant data.
This information on Android apps from different datasets complements each other.
New insights could be gained from combining datasets and drawing connections between the existing data points.
Future work could facilitate this kind of research by creating a meta-dataset which links data on Android applications in existing datasets.

\section{Conclusions}\label{sec:related:conclusion}

Researchers of Android applications have a vast amount of data at hand.
There are already many datasets containing executable artifacts.
App store metadata is plentiful and public albeit difficult to access.
Many studies report this problem, especially in accessing data from \play.
However, insight into source code is limited because the vast majority of apps is proprietary.
Several studies tried to gather and combine source code with other app metadata.

Datasets of app store data and executables have the advantage, that they are independent of licensing of the application source code.
Data from marketplaces can be scraped for free while APK archives can be downloaded from app stores.
On the other hand, source code for proprietary applications is to a large extent not available at all.
Having both a comprehensive dataset of (almost) all available apps -- as with AndroZoo -- and having access to source code is unfortunately not reconcilable.

\appendix

\newgeometry{margin=1cm}
\begin{landscape}
    \section{Survey results}\label{appendix:results}

    \scriptsize
    \pagestyle{empty}

    \renewcommand*{\arraystretch}{1.4}

    \begin{longtable}{>{\itshape}R{19mm} l R{21mm} R{2cm} l l l l l l R{2cm} R{21mm}}
         & Year & Summary & Data gathered & Number of apps & Source code & Commit history & Executables & Google Play link & F-Droid link & Sourced from & Remarks
        \\ \hline\hline
        \endhead
        Appannie & & Commercial data aggregation & ongoing & 14+ million & no & no & no & yes & no & ``All major app stores'' & \url{https://appannie.com}
        \\\hline
        AppBrain & & Commercial app meta data database (Play mirror) & ongoing & 3,749,507 & no & no & no & yes & no & Google Play & \url{https://appbrain.com}
        \\\hline
        AppZoom & & Commercial app review and analysis & ongoing & ? & no & no & no & yes & no & Google Play, Apple Store & \url{https://appszoom.com}
        \\\hline
        Zhou and Jiang~\cite{zhou_dissecting_2012} & 2012 & Selection of malware. & 2010 to 2011 & 1,260 & no & no & yes & no & no & Security announcments, publications from anti virus vendors and researchers. & Stopped sharing their data in 2015.
        \\\hline
        Aafer \etal~\cite{aafer_droidapiminer:_2013} & 2013 & Static analysis of APKs & July 2012 & around 20,000 & no & no & yes & partially & no & McAffee, \cite{zhou_dissecting_2012}, Google Play &
        \\\hline
        Minelli and Lanza~\cite{minelli_software_2013} & 2013 & Static analysis of source code & 2013 (?) & 20 & yes & yes & no & yes & yes & F-Droid & \url{http://samoa.inf.usi.ch}
        \\\hline
        Petsas \etal~\cite{petsas_rise_2013} & 2013 & Monitoring of metrics on app stores. & Mar -- Aug 2012 & 300,000 & no & no & no & no & no & SlideMe, 1Mobile, AppChina, Anzhi & Sources selected for accurate number of downloads reported.
        \\\hline
        Zheng \etal~\cite{zheng_droid_2013} & 2013 & Signature based analytics & ongoing & 150,368 & no & no & yes & no & no & Google Play and other app stores, malware forums. &
        \\\hline
        Arp \etal~\cite{arp_drebin:_2014} & 2014 & Vector based analytics &  Aug 2010 -- Oct 2012 & 123,453 & no & no & yes & partially & no & Google Play, Chinese and Russion app stores, malware forums, \cite{zhou_dissecting_2012} &
        \\\hline
        Gorla \etal~\cite{gorla_checking_2014} & 2014 & Signature based anomaly detection & Winter and Spring 2013 & 32,136 & no & no & yes & yes & no & Google Play &
        \\\hline
        Linares-V\'asques \etal~\cite{linares-vasquez_mining_2014} & 2014 & Detect energy optimizations from usage patterns & 2014 (?) & 55 & no & no & no & yes & no & Google Play & \url{http://www.cs.wm.edu/semeru/data/MSR14-android-energy}
        \\\hline
        Lindorfer \etal~\cite{lindorfer_andrubis1000000_2014} & 2014 & Automated dynamic and static analysis & 2012 -- 2015 & 1,034,999 & no & no & yes & yes & no & submissions, malware feeds & discontinued
        \\\hline
        Moonsamy \etal~\cite{moonsamy_mining_2014} & 2014 & Fingerprinting of permissions & Aug 2010 -- Oct 2011 & 1,227 & no & no & yes & no & no & SlideME, Pandaapp & Used \cite{zhou_dissecting_2012} as complementary set of malware.
        \\\hline
        Corral and Fronza~\cite{corral_better_2015} & 2015 & Relating source code quality to market success & 2013 & 100 & yes & no & no & yes & yes & F-Droid &
        \\\hline
        Freiling \etal~\cite{freiling_empirical_2015} & 2015 & Evaluation of obfuscation transformations & 2015 (?) & 240 & no & no & yes & no & yes & F-Droid &
        \\\hline
        Krutz \etal~\cite{krutz_dataset_2015} & 2015 & Collection and static analysis of open source Android applications with metadata and commit history & 2015 (?) & 1,179 & yes & yes & yes & no & yes & F-Droid &  Open Source only. Website hosting dataset seems to be defunct.
        \\\hline
        Lamba \etal~\cite{lamba_pravaaha:_2015} & 2015 & Static analysis on source code & July 2014 & 1,120 & yes & no & no & no & yes & F-Droid & Extensive description of F-Droid
        \\\hline
        Linares-V\'asques \etal~\cite{linares-vasquez_how_2015} & 2015 & Survey among developers on performance issues & 2015 (?) & 485 & yes & yes & no & no & no & GitHub & Identify Android apps in Github repositories by manifest file.
        \\\hline
        Malavolta \etal~\cite{malavolta_hybrid_2015} & 2015 & Study of users' perception of hybrid apps & Nov 2014 & 11,917 & no & no & no & yes & no & Google Play &
        \\\hline
        Tufano \etal~\cite{tufano_when_2015} & 2015 & Identify bad programming practices from commit history & 2015 (?) & 70 & yes & yes & no & no & yes & F-Droid & Next to Android apps, also Apache and Eclipse projects are studied.
        \\\hline
        Allix \etal~\cite{allix_androzoo:_2016} & 2016 & Collection of APKs for analysis. & ongoing & 5,842,525 &	no &	no &	yes
        & yes\footnote{
            Link can be constructed from package name if available on Google Play.
            \label{foot:link-to-play}
        }
        & yes \footnote{
            Link can be constructed from package name if available on F-Droid.
            \label{foot:link-to-fdroid}
        }
        & Various app markets, Torrents, \cite{zhou_dissecting_2012}. & The collection is still growing. Apps are labeled with the markets they are found on.
        \\\hline
        Avdiienko \etal~\cite{avdiienko_calappa:_2016} & 2016 & Scraping tool to combine data about Android apps from various sources & --- & ---
        & yes\footnote{
            Depending on crawler module and source
            \label{foot:calappa:qualification}
        } & yes\footref{foot:calappa:qualification} & yes\footref{foot:calappa:qualification} & yes\footref{foot:calappa:qualification} & yes\footref{foot:calappa:qualification} & Google Play, apkmirror.com, F-Droid &
        \\\hline
        Bao \etal~\cite{bao_how_2016} & 2016 & Identify power management activities from Git commits & 2016 (?) & 1,273 & yes & yes & no & no & yes & F-Droid, \cite{krutz_dataset_2015} &
        \\\hline
        Das \etal~\cite{das_quantitative_2016} & 2016 & Study of performance related commits	& 2016 (?) & 2,443 &	yes &	yes & no & yes & yes & F-Droid, Wikipedia, Github README files &
        \\\hline
        McIlroy \etal~\cite{mcilroy_fresh_2016} & 2016 & Study of update frequency of apps & 2014 & 10,713 & no & no & no & yes & no & Google Play &
        \\\hline
        Nayebi \etal~\cite{nayebi_analysis_2016} & 2016 & Analysis of app release cycles & 2016 (?) & 1,844 & yes &  &  & yes & yes & F-Droid &
        \\\hline
        Grano \etal~\cite{grano_android_2017} & 2017 & Tracking of user feedback from reviews to changes & 2017 (?) & 395 & yes & no & yes & yes & yes & F-Droid, Google Play & Includes 297,323 reviews
        \\\hline
        Krutz \etal~\cite{krutz_who_2017} & 2017 & Static analysis of app permissions of apps in F-Droid. & 2017 (?) & 1,402 & yes & yes & no & yes & yes & F-Droid, GitHub &
        \\\hline
        Meng etal~\cite{meng_androvault:_2017} & 2017 & Knowledge graph from results of static and dynamic analysis & since 2013 & $>$ 5,000,000 & no & no & yes & partially & partially & 28 app stores including Google Play and F-Droid &
        \\\hline
        Stojkovski~\cite{stojkovski_thresholds_2017} & 2017 & Thesis on various software metrics for Android apps  & 2014 -- 2017 (?) & 865 & yes & yes & no & no & yes & F-Droid & Did not source from Sourceforge for lack of scalable access to Android apps
        \\\hline
        Geiger \etal~\cite{geiger_graph-based_2018} & 2018 & Graph database combining metadata on Google Play and GitHub with commit history & 2017 -- 2018 & 8,431 & yes & yes & no & yes & no & GitHub, Google Play &
    \end{longtable}
\end{landscape}
\restoregeometry

\addcontentsline{toc}{section}{References}
\bibliographystyle{plain}
\bibliography{bib/android_datasets_2018} 

\end{document}